\documentclass[twocolumn,showpacs,preprintnumbers,amsmath,amssymb]{revtex4}

\usepackage{graphicx}
\usepackage{dcolumn}
\usepackage{bm}
\usepackage{float}
\usepackage{natbib}

\newcommand{\ket}[1]{\vert #1 \rangle}

\newcommand{\ef}{\hat{\mathcal{E}}}

\begin{document}

\title{Configurable unitary transformations and linear logic gates using quantum memories}
\author{G. T. Campbell$^1$, O. Pinel$^1$, M. Hosseini$^1$, T. C. Ralph$^2$ and B. C. Buchler$^1$ and P. K. Lam$^1$}
\affiliation{$^1$Centre for Quantum Computation and Communication Technology, Department of Quantum Science, The Australian National University, Canberra, Australia}
\affiliation{$^2$Centre for Quantum Computation and Communication Technology, School of Mathematics and Physics, University of Queensland, Brisbane, Queensland 4072, Australia}

\date{\today}

\begin{abstract}

We show that a set of optical memories can act as a configurable linear optical network operating on frequency-multiplexed optical states. Our protocol is applicable to any quantum memories that employ off-resonant Raman transitions to store optical information in atomic spins. In addition to the configurability, the protocol also offers favourable scaling with an increasing number of modes where N memories can be configured to implement an arbitrary N-mode unitary operations during storage and readout. We demonstrate the versatility of this protocol by showing an example where cascaded memories are used to implement a conditional CZ gate. 
\end{abstract}

\maketitle

One of the key elements in an optical network for transmitting and manipulating quantum information is an optical memory that is capable of on-demand storing and recalling of optical states without significant loss or addition of noise \cite{Lvovsky:NPhoton:2009}. In addition to acting as a tool for synchronising elements of an optical network, a quantum memory is crucial for the implementation of linear optical quantum computation, which relies on the ability to feed-forward measurement results to the configuration of linear logic gates at a later point in an optical network \cite{Knill:2001,Kok:2007}.

The need for both quantum memories and scalable linear optical networks has driven advances in both areas. High-efficiency and high-fidelity quantum memories have been developed \cite{Hosseini:Nphys:2011,Afzelius:2010,Nunn:2010:NPhoton:4,Tittel:2010:lpor:4,Appel:2008,Figueroa:2011}; integrated and configurable photonic circuits that can perform operations on path-encoded quantum channels have been demonstrated \cite{Peruzzo:2007}. Here, we propose a novel architecture for manipulating optical quantum information by using quantum memories, not just for the synchronisation and buffering of optical states, but also to perform linear operations on them.

The underlying system for our proposal is a quantum memory in which an optical state, $\ef_{\rm in}$, is mapped, via a far-off-resonant Raman transition, to a collective spin-wave, $\hat{S}$, in an ensemble of $\Lambda$-type three-level atoms \cite{Gorshkov:2007:PRA:76}. The temporal mode of $\ef_{\rm in}$ and the spatial mode of $\hat{S}$ will depend on the particular method of storage; however, we assume that $\ef_{\rm in}$ is tailored such that the mapping $\ef_{\rm in} \rightarrow \hat{S}$ is efficient and reversible. Candidate storage methods include using appropriate temporal shaping of a strong coupling field \cite{Nunn:2007:PRA:75,Nunn:2010:NPhoton:4,Moiseev:2013,Moiseev:2011} or a reversible distribution of two-photon detunings \cite{Longdell:2008p8530,Hetet:2008p5840, PhysRevA.80.012320}. Configurable beam-splitting into temporal modes has been demonstrated for both of these storage methods \cite{PhysRevLett.108.263602, Hosseini:2009p8466}; we show that extending these memories to operate on modes that are separated in frequency allows more general operations.

\begin{figure*}
\centerline{\includegraphics[width=2 \columnwidth]{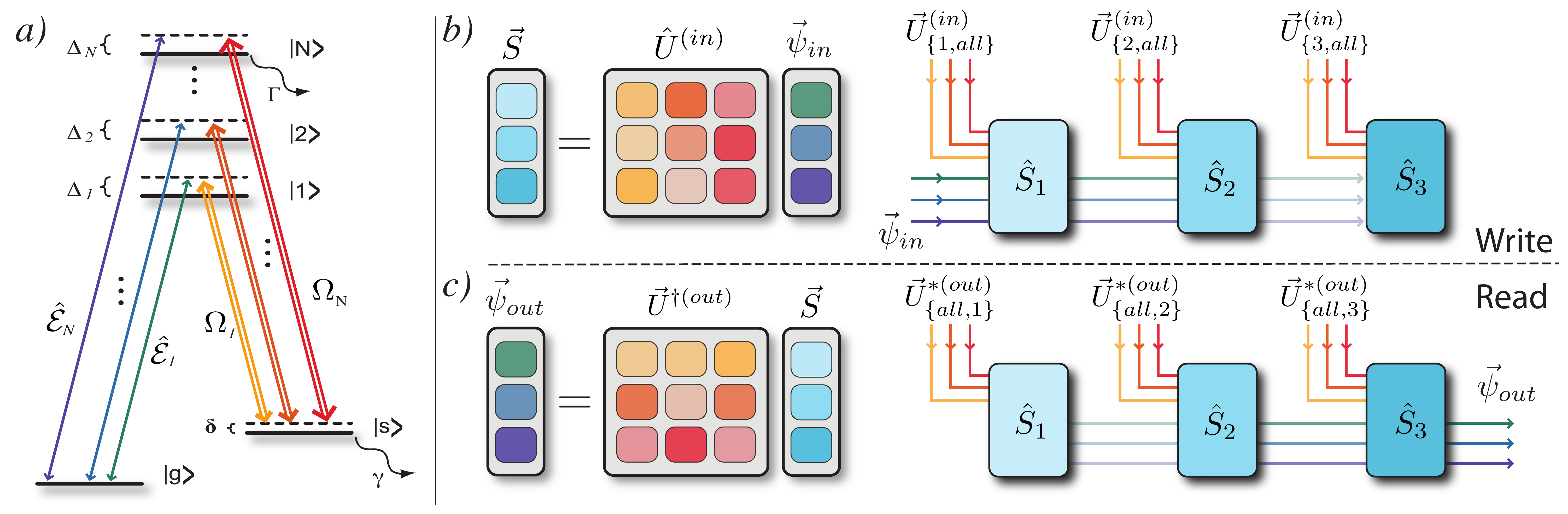}}
 \caption{
    a) The multi-$\Lambda$ level structure used in the protocol consists of a ground-state $\ket{g}$, a meta-stable state $\ket{s}$ and $N$ excited states, $\ket{e_k}$. Optical information carried in the probe modes, $\ef_k$, can be mapped onto the spin-coherence between $\ket{g}$ and $\ket{s}$ via a two-photon Raman scattering mediated by the coupling fields, $\Omega_k$. b) Controlling the coupling field amplitudes results in the mapping of a unique superposition of the probe modes into each of a set of memories that are placed in series along the optical path. By selecting appropriate coupling field amplitudes $\vec{U}^{\rm (in)}_{\{i, \rm{all}\} } $ (represented by column vectors), an arbitrary unitary operation $\hat{U}^{\rm (in)} = [\vec{U}^{\rm (in)}_{\{ 1, \rm{all} \} }, \vec{U}^{\rm (in)}_{\{ 2, \rm{all} \} } , \vec{U}^{\rm in}_{ \{ 3, \rm{all} \} } ]^{T} $ is implemented during the mapping from frequency-separated optical modes to spatially separated spin-waves. c) A second arbitrary unitary operation $\vec{U}^{\dagger (\rm out)} = [\vec{U}^{\ast \rm (out)}_{\{ \rm{all}, 1 \} }, \vec{U}^{\ast \rm (out)}_{\{ \rm{all}, 2 \} } , \vec{U}^{\ast \rm (out)}_{ \{  \rm{all},3 \} } ]$ can applied during the read phase of the scheme.
 }
 \label{implementation}
\end{figure*}

We consider a three-level system that includes multiple signal fields which are separated in frequency. The atomic structure, depicted in Fig.~\ref{implementation}(a), consists of a ground state, $\ket{g}$, a meta-stable state, $\ket{s}$, with dephasing rate $\delta$ and $N$ excited states, $\lbrace \ket{e_1}, \ldots, \ket{e_N} \rbrace$. Each of the signal fields, $\ef_k$, is coupled to the spin excitation by a corresponding bright coupling field $\Omega_k$ at a detuning $\Delta_k$, resulting in multiple Raman-$\Lambda$ transitions that drive the spin excitation of the atomic ensemble. Each excited state is assumed to have the same decay rate, $\Gamma$, and each of the Raman transitions the same two-photon detuning, $\delta$.

In the far-detuned regime, $\Delta_k \gg \beta \Gamma$ where $\beta$ is the optical depth \cite{Gorshkov:2007:PRA:76}, we can adiabatically eliminate the excited states by assuming that $\partial_t\hat{\sigma}_{gk} \approx 0$. Making dipole, rotating-wave and pure-state approximations, the Heisenberg equations of motion for the probe field and collective atomic operators \cite{Hetet:2008p5840} can then be written
\begin{subequations}\label{TMB}
	\begin{align}
{\partial_{t'}\hat\sigma}_{gs} &= - (\gamma' + i \delta') \hat{\sigma}_{gs} + i g \sum\limits_{k=1} \frac{\Omega^*_k}{\Delta_k} \ef_k, \label{TMB:1} \\
\partial_{z'} \ef_k &= i \mathcal{N} \frac{\Omega_k}{\Delta_k}\hat{\sigma}_{gs}, \label{TMB:2}
	\end{align}
\end{subequations}
where $\lbrace{z' = z, t' = t - z/c\rbrace}$ is a moving reference frame, $g$ is the light-atom coupling and $\mathcal{N}$ is the effective linear atomic density. The two-photon detuning, $\delta' = \delta - \sum_k (\vert \Omega_k \vert/\Delta_k)$, and dephasing rate of $\ket{s}$, $\gamma' = \gamma + \Gamma\sum_k (\vert \Omega_k \vert/\Delta_k)^2$, have been adjusted to account for the  ac-Stark shift and power broadening, respectively. The dispersion for each field is incorporated into the envelope functions with the transformations  $\ef_k \rightarrow \ef_k e^{-i g \mathcal{N} z /\Delta_k}$, and $\Omega_k \rightarrow \Omega_k e^{-i g \mathcal{N} z /\Delta_k}$ prior to transforming to a moving reference frame. Each signal will acquire a phase of $\beta \Gamma/\Delta_k \ll 1$, resulting in a phase mismatch if the detunings for each $\Lambda$ system differ. This phase mismatch is small in the far-detuned regime, but could also be eliminated with the introduction of a small angle.

 The protocol is based on the observation that the second term on the right side of eq.~(\ref{TMB:1}) describes the projection of the input modes onto the vector defined by the amplitudes of $\Omega_k/\Delta_k$. We can define a transformed set of modes $\ef'_j = \sum_{k=1}^{N} U_{jk} \ef_k$ that includes the mode $\ef_c'$, which is in turn defined by
 \begin{subequations}
	\begin{align}\label{OmegaDef}
		\ef'_c &\equiv \frac{1}{\tilde{\Omega}}\sum\limits_{k=1}^{N} \frac{\Omega^*_k}{\Delta_k} \ef_k\\
		\tilde{\Omega} &= \sqrt{\sum_{k=1}^{N} \vert\frac{\Omega_k}{\Delta_k}\vert^2}.
	\end{align}
\end{subequations}
In this transformed basis, equation ($\ref{TMB}$) reduces to 
\begin{subequations}\label{SMB}
	\begin{align}
{\partial_{t'}\hat\sigma}_{gs} &= - (\gamma' + i \delta') \hat{\sigma}_{gs} + i g \tilde{\Omega} \ef'_c, \label{SMB:1} \\
\partial_{z'} \ef'_c &= i \mathcal{N} \tilde{\Omega}\hat{\sigma}_{gs}, \label{SMB:2} \\
\partial_{z'} \ef'_{k \neq c} &= 0, \label{SMB:3}
	\end{align}
\end{subequations}
which describes a single mode, $\ef'_c$, interacting with an ensemble of two-level atoms with detuning $\delta'$, linewidth $\gamma'$ and effective coupling strength $g \tilde{\Omega}$. The remaining modes, $\ef'_{k \neq c}$ propagate unimpeded by the atomic ensemble.

Storage of $\ef'_c$ results in the mapping $\sum_{k=1}^{N} U_{ck} \ef_k \rightarrow \hat{S}$, where $\hat{S}$ is a collective spin excitation of $\hat{\sigma}_{gs}$. The application of an arbitrary unitary operation can be accomplished by placing $N$ memories in series along the optical path. The mapping from the input optical state to the memories, illustrated in Fig.~(\ref{implementation}b), will be $\sum_{k=1}^{N} U^{\rm (in)}_{jk} \ef_k \rightarrow \hat{S_j}$ if the coupling field amplitudes in the $j$-th memory are set such that
 \begin{equation}\label{UDef}
U^{\rm (in)}_{jk} = \frac{1}{\tilde{\Omega}} \frac{\Omega^*_k}{\Delta_k}.
\end{equation}
Similarly, on recall, the mapping $\sum_{j=1}^{N}U^{\ast \rm (out)}_{kj}\hat{S_j} \rightarrow  \ef_k$ is performed, as illustrated in Fig.~(\ref{implementation}c), with the coupling fields defined as in eq.~(\ref{UDef}). This allows two independent unitary operations to be performed: one during the write phase, with the resultant state being stored across the set of memories, and one during the read phase, with the result of the operation being restored to a frequency-multiplexed optical state.

The system scales favourably with an increasing number of modes: it requires N memories for a rank-N operation. The protocol does require a total $N^2$ optical coupling fields; however, at large overall detunings the coupling fields in each memory copropagate and could be generated by a single electro-optical element. Furthermore, the total power of the coupling fields in each memory is independent of the number of modes. Efficient storage of $\ef_c'$ in eq. \ref{SMB} requires a large effective optical depth of the Raman transition, $\beta_{\rm eff} = \beta \tilde{\Omega}^2 \Gamma \gamma^{-1} $, where $\beta = g \mathcal{N} L/\Gamma$ is the resonant optical depth. The effective transition strength depends on the {\it total} optical power, scaled to the inverse detuning, in all of the coupling fields. An additional advantage is that loss due to memory inefficiency is incurred only once for each mode and, provided that absorption due to the far-detuned excited state is negligible, the resulting overall efficiency is independent of the number of modes.

The requirement that each $\Lambda$ transition correspond to a separate excited state is likely prohibitive of a practical implementation of the protocol. It is therefore of interest to determine under what conditions it is possible to use multiple $\Lambda$ transitions that are detuned from a single excited state. In the limit where the transitions are well separated, each $\Lambda$ system can be adiabatically eliminated independently (see supplementary material) and we recover eq.~(\ref{TMB}). The condition for this to be valid is that $\hat{\sigma}_{gs}$ must evolve slowly with respect to the frequency separation of the modes so we examine how it is influenced by interactions between the $\Lambda$ systems.

We proceed by adiabatically eliminating the excited state under the assumption that the detuning is the fastest timescale of the system: $\beta \Gamma  \ll \Delta$ and $\vert \Delta_k - \Delta \vert  \ll \Delta$, where the detunings are centred around a mean frequency, $\Delta$. In this  limit, the equations of motion for a probe field propagating through an ensemble of three-level atoms are:
\begin{subequations}\label{SMB_OES_G}
	\begin{align}
	  {\partial_{t'}\hat\sigma}_{gs} &= -\left(\gamma + i \delta + i  (\Delta - i \Gamma)(\vert \Omega\vert^2/\Delta^{2})\right) \hat{\sigma}_{gs} \nonumber\\
	  & \qquad {} + i g (\vert\Omega\vert/\Delta) \ef_c , \label{SMB_OES_G:1} \\
\partial_{z'} \ef_c &= i \mathcal{N} (\vert\Omega\vert/\Delta) \hat{\sigma}_{gs}, \label{SMB_OES_G:2}
	\end{align}
\end{subequations}
where a $\ef_c$ is a unitary transformation of the probe field, $\ef_c \equiv \Omega^* \vert \Omega \vert^{-1} \ef$. The coupling field, $\Omega = \sum_k \Omega_k \exp[i (\Delta_k - \Delta) t']$, and the probe field, $\ef_c$, contain rapidly oscillating components.
We solve equation (\ref{SMB_OES_G}) and examine the contribution of the rapidly varying terms.

For simplicity, we first solve eq.~\ref{SMB_OES_G:1} in the absence of a probe field with the result:
\begin{subequations}\label{Sol1}
  \begin{align}
    \hat{\sigma}_{gs} &= \alpha e^{-(\gamma' + i \delta') t'} \prod\limits_{j,k \neq j} \exp\left(-i \frac{\Omega^*_k \Omega_j}{ \delta_{k,j}} \frac{\Gamma - i \Delta}{\Delta^2} e^{i \delta_{k,j}t'}\right), \label{Sol1:1}\\
                                &\approx \alpha e^{-(\gamma' + i \delta') t'} \left( 1 - \sum\limits_{j,k \neq j} \frac{\Omega_k^*\Omega_j}{\Delta \delta_{k,j}} e^{i \delta_{k,j} t}\right), \label{Sol1:2}
  \end{align}
\end{subequations}
where $\alpha$ is an integration constant, $\delta_{k,j} \equiv ( \Delta_k - \Delta_j )$, and we have assumed that $\Delta \gg \Gamma$. To move from (\ref{Sol1:1}) to (\ref{Sol1:2}) we have assumed that $\vert \delta_{k,j} \vert \gg \vert \frac{\Omega^*_k \Omega_j}{\Delta} \vert$ and eliminated higher-order terms. The magnitude of the perturbation will depend on the particular amplitudes of the fields. We can place a conservative requirement on the magnitude of $\delta_{j,k}$ by assuming equal spacings between modes and roughly equal distribution of the coupling field power across the mode spectrum. With these conditions, we require 
\begin{equation}\label{Ineq1}
\vert \delta_{j,k} \vert \gg \frac{\Delta \vert \tilde{\Omega} \vert^2}{\sqrt{N}}.
\end{equation}
The $\sqrt{N}$ is a result of assuming that the amplitudes of the terms in sum do not interfere constructively and can be added in quadrature.

With the oscillating component of $\vert \Omega (t') \vert^2$ omitted, we now consider a general solution of (\ref{SMB_OES_G:1}) that includes the probe field driving terms:
  \begin{align}
    \hat{\sigma}_{gs} = &\alpha e^{-(\gamma' + i \delta') t'} + \frac{i g}{\gamma' + i \delta'}\sum\limits_{k} \frac{\Omega^*_k }{\Delta}\ef_k \nonumber \\
      &+ i g \sum\limits_{k,j \neq k} \frac{\Omega^*_j }{\Delta}\ef_k \frac{e^{i \delta_{k,j} t'}}{\gamma' + i(\delta' + \delta_{k,j})},
  \end{align}      
where we have assumed a form of $\ef = \sum\limits_k \ef_k e^{i (\Delta_k - \Delta) t'}$ for the probe field.  Again, the magnitude of the rapidly oscillating term in $\hat{\sigma}_{gs}$ will depend on the amplitudes of $\lbrace \Omega_k^* \ef_k \rbrace$. A conservative restriction would be
\begin{equation}\label{Ineq2}
\vert \delta_{k,j} \vert \gg \sqrt{N} \times \lbrace \gamma', \vert \delta' \vert \rbrace. 
\end{equation}
This requires that the frequency separation of the modes be large compared to the memory bandwidth.

To place estimates on the parameters for a viable implementation, we consider the memory presented in \cite{Sparkes:2013}, which uses a gradient echo memory as the storage mechanism. This type of memory has demonstrated noiseless and efficient operation \cite{Hosseini:Nphys:2011,Hosseini:NComm:2011}. An implementation that uses a cold atomic vapour, such as \cite{Sparkes:2013}, is advantageous because the coupling fields in each memory do not need to copropagate with the signal fields but can be offset by a small angle. This allows a single memory to be subdivided into segments, each of which is addressed by its own set of coupling fields, without the need for beamsplitters between the segments.

The relevant quantites from \cite{Sparkes:2013} are: a maximum available optical depth of 1000 in a 5 mm long ensemble, a 10 $\mu$s pulse duration, a detuning of 250 MHz and a coupling field power of 350 $\mu$W. Dividing the memory into segments, each with an optical depth of 100, would allow a ten mode implementation of the protocol with modes separated by 50 MHz which satisfies (\ref{Ineq2}), the more stringent of the two inequalities, by a factor of $167$. This would require a total coupling field power of 32 mW and, in principle, would be $\approx 95\%$ efficient. The total bandwidth, 500 MHz, is less than the bandwidth of available electro-optic modulators, meaning that all of the coupling fields could be generated with only ten electro-optic elements. Dispersion between signal fields in this example would cause a $\approx\lambda/4$ difference between the nearest and furthest detuned modes, requiring a $1.6^{\circ}$ spread in the coupling field angles to correct. Increasing the base detuning to 750 MHz, however, would reduce the phase error due to dispersion to $\approx\lambda/20$ without any angle between the coupling fields in each segment.

Numerical simulations \cite{XMDS,Hetet:2008p7696} were performed for similar parameters to verify the performance of the protocol, although a mode spacing of 15 MHz rather than 50 MHz was used to reduce computation time. The results demonstrated an overall efficiency of ($91.2\pm0.2$)\% with an overlap of ($0.988 \pm 0.003$) between the simulated and ideal outputs after two operations.

\begin{figure*}
\centerline{\includegraphics[width= 2 \columnwidth]{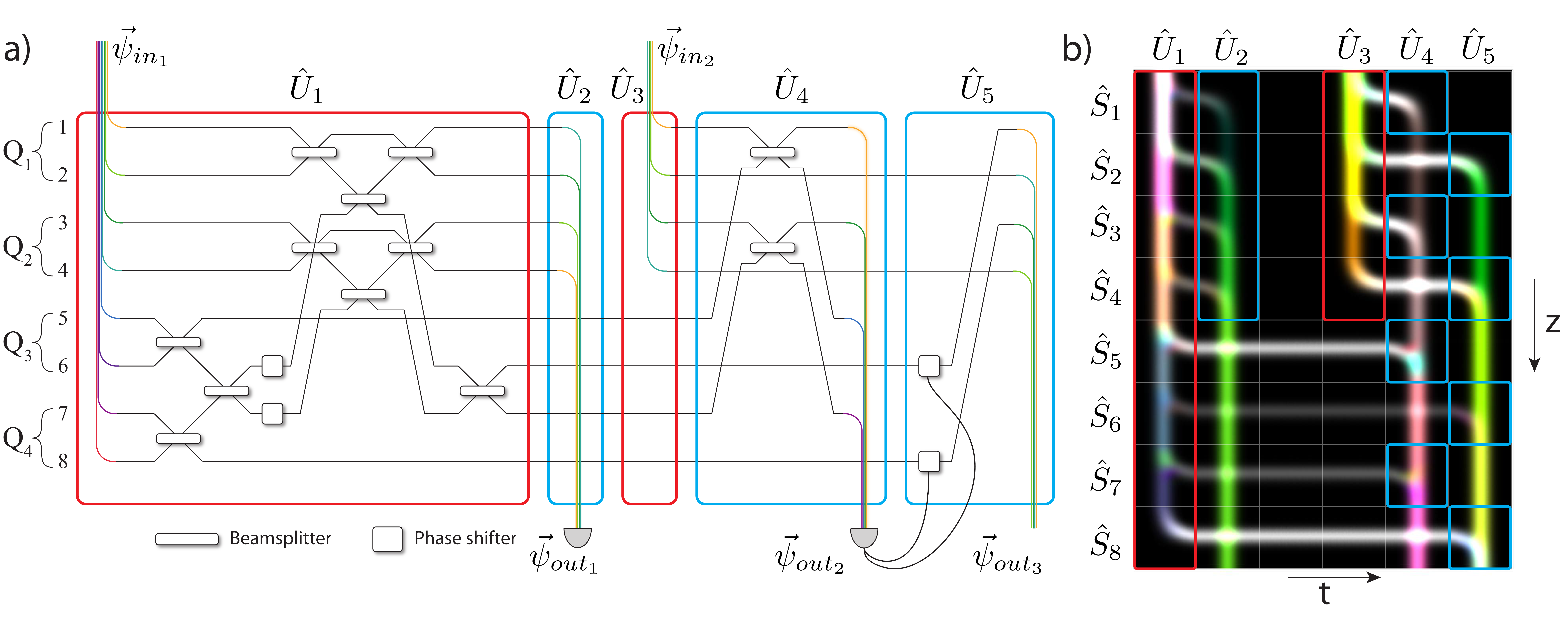}}
 \caption{
     A numerical simulation of the protocol implementing the unitary operation required to perform a conditional c-z gate as described in \cite{Knill:2001}. (a) The network of linear optical elements as originally proposed. The qubits are path-encoded photons with $\hat{Q}_n$ representing the $n$th qubit. In this diagram, photons are injected from the top ($\hat{\psi}_{in_n}$) and read out from the bottom ($\hat{\psi}_{out_n}$) for consistency with the simulation. (b) The simulation results, illustrating the propagation of fields through the memory elements as each unitary operation is performed. Each optical frequency is represented by a different color and atomic spin waves are represented as white. The optical modes co-propagate while the spin waves remain spatially separated, each horizontal segment, labeled $\hat{S}_1$ through $\hat{S}_N$, representing a memory element. Each vertical segment is a temporal window in which a storage or recall event can occur to map information between the (frequency-encoded) optical basis and the (path-encoded) spin-wave basis. Every area in the grid therefore has an independently controllable set of coupling field amplitudes. The optical fields enter from the top and propagate rapidly along the spatial dimension until they are stored in a memory cell. On recall, they continue to propagate and exit at the bottom of the grid. The areas outlined in red are storage operations and the areas outlined in blue are recall operations.
 }
 \label{xmds_fig}
\end{figure*} 

A ten-mode implementation of the protocol would be sufficient to perform non-trivial linear gates. As a demonstration of the versatility, we simulated an operation that is useful to linear optical quantum computing; the implementation of a conditional cz gate as described in Ref.~\cite{Knill:2001}. Figure \ref{xmds_fig} illustrates how this gate can be performed on a pair of dual-rail qubits using conditional measurements and feeding-forward of measurement results. The left side of the figure (a) shows the network of one- and two-mode linear operations as originally proposed. The right side (b) shows the same unitary operations as implemented in our protocol using a set of eight memories.

We refer the reader to \cite{Knill:2001,Kok:2007} for a description of the linear network shown in \ref{xmds_fig} (a) and focus only on how the network can be translated into the memory-based protocol. In our proposed implementation, four memories are conditionally prepared by the operation $\hat{U}_{1}$, acting during storage, and a detection of the state of the other four modes, performed by a read-out step, $\hat{U}_{2}$. If the operation has been successful, the state of interest is then written into the memories and the successful gate result teleported onto modes of interest via another linear operation, $\hat{U}_{4}$, and a conditional phase-shift of the remaining qubits, $\hat{U}_{5}$.

This example highlights some of the advantages of this protocol. The combination of an optically configurable operation with integrated memory provides a simple path to measurement-based non-linearities. Operations involving a large number of modes can be performed in a single step with the result being stored in memory for use at a later point in a computation. The protocol is spatially single-mode, providing inherent interferometric stability and only one type of element is required, eliminating interfaces between devices.  The major drawback, replacing simple optical elements with atomic ensemble-based memories, is mitigated by the fact that networks of linear optics would likely require similar memories for synchronisation and to implement the feeding-forward of measurement results.

Integrating the ability to perform arbitrary unitary transformations into a quantum memory provides a potential platform for manipulating quantum states of light. In particular, the operation on frequency multiplexed states makes the protocol naturally suited to work with optical parametric oscillator sources that produce a frequency comb of two-mode squeezed states \cite{PhysRevLett.101.130501,PhysRevLett.107.030505}. Furthermore, the underlying mechanism for the operation has already been demonstrated in the case of electromagnetically induced transparency \cite{Campbell:2009p3114, Appel:2007p10361} and for the case of two modes in a gradient echo memory \cite{Campbell:2012p033022}. The recent development of a Raman scattering based optical memory in a solid-state ensemble \cite{Goldschmidt:13} is an encouraging step towards compact memories that are suitable for implementation of the protocol.

This research was conducted by the Australian Research Council Centre of Excellence for Quantum Computation and Communication Technology (project no. CE110001027).

\bibliographystyle{unsrt}
\bibliography{bibs}

\end{document}